\documentclass{aa}
\usepackage{graphicx}
\usepackage{natbib}

\newcommand{\centercell}[1]{\multicolumn{1}{c}{#1}}

\begin{document}
\title{Image reduction pipeline for the detection of variable sources
in highly crowded fields}
\author{C.A.~G\"ossl \and A.~Riffeser}
\authorrunning{G\"ossl \& Riffeser}
\offprints{C.A.~G\"ossl}
\mail{cag@usm.uni-muenchen.de}
\institute{Universit\"ats-Sternwarte M\"unchen, Scheinerstr. 1, 
  D-81679 M\"unchen, Germany\\
  email: cag@usm.uni-muenchen.de}
\date{Received August 6, 2001; accepted October 30, 2001}
\abstract{
We present a reduction pipeline for CCD ({\em charge-coupled
device}\/) images which was built to
search for variable sources in highly crowded fields
like the M\,31 bulge and to
handle extensive databases due to large time series.
We describe all steps of the standard reduction in detail with
emphasis on the realisation of per pixel error propagation:
Bias correction, treatment of bad pixels, flatfielding, 
and filtering of cosmic rays.
The problems of conservation of PSF ({\em point spread function})
and error propagation in our image alignment procedure
as well as the detection algorithm for variable sources are discussed:
We build difference images via image convolution with a technique
called OIS
\citep[{\em optimal image subtraction}, ][]{1998ApJ...503..325A},
proceed with an automatic detection of variable sources in noise
dominated images and finally apply a PSF-fitting, relative
photo\-metry to the sources found.
For the WeCAPP project
\citep{2001A&A...0000..00R}
we achieve $3\sigma$ detections for variable sources
with an apparent brightness of e.g.
$m = 24.9\;\mathrm{mag}$ at their minimum and a variation of
$\Delta m = 2.4\;\mathrm{mag}$ (or
$m = 21.9\;\mathrm{mag}$ brightness minimum and a variation of
$\Delta m = 0.6\;\mathrm{mag}$)
on a background signal of $18.1\;\mathrm{mag}/\mathrm{arcsec}^2$ based
on a $500\;\mathrm{s}$ exposure with $1.5\;\mathrm{arcsec}$ seeing at
a $1.2\;\mathrm{m}$ telescope.
The complete per pixel error propagation allows us to give accurate
errors for each measurement.
\keywords{Methods: data analysis -- Methods: observational --
Techniques: image processing -- Techniques: error propagation --
Techniques: optimal image subtraction}
}
\maketitle
\section{Introduction}
Astronomical imaging in optical wavebands is performed nearly
exclusively with {\em charge-coupled devices}\footnote{
The history of CCDs in astronomy and a basic description of them can
be found in
\citet{1997wiley.book.....M,1991cacu.book.....B,1990ccda.proc.....J,1986ARA&A..24..255M}.}
today.
Despite the numerous advantages of modern CCDs,
their images still have to be corrected for a couple of
disturbing influences and effects before one can base advance in
science on them.
Here, we will focus on the problems arising with optical, ground based
imaging and time-series observations to find and measure variable
sources either hidden in a bright background
(e.g. a variable star in its host galaxy) or a crowded field or even
in a combination of both.

The search for variable objects with common photometry methods
becomes very ineffective in crowded fields because of blending.
\citet{1995adass...4..297P}
show algorithms for registering, matching the point spread functions
(PSFs), and matching the intensity scales of two or more images in
order to detect transient events.
\citet{1996AJ....112.2872T}
propose a method called {\em Difference Image Analysis} (DIA)
where the point spread function (PSF),
describing the projection of a point source onto the image plane,
is matched by calculating a convolution kernel in Fourier space.
The application to real data has been shown for Galactic Microlensing
\citep{1999ApJS..124..171A}
as well as for microlensing in M\,31 \citep{1999gady.conf..409C}.
A new method for {\em Optimal Image Subtraction} (OIS) of two images
has been designed by \citet{1998ApJ...503..325A}.
They derive an optimal kernel solution from a simple least-squares
analysis using all pixels of both images.
This method has been used successfully by different projects
(\citealp[OGLE,][]{2000AAS...197.6702W};
\citealp[MOA,][]{2001MNRAS.327..868B};
\citealp[DIRECT,][]{2001AJ....121.2032M};
etc.).

\begin{table*}
\centering
\small{
\begin{tabular}{p{0.12\textwidth}p{0.83\textwidth}}
\hline
\hline
origin & property\\
\hline
Detector: &
Photosensitive area and additional ``borders''
(prescan, postscan, overscan),
geometric variation of pixel size and edge pixels,
pixel sensitivity variations and pixel defects (cold pixel, hot pixel,
trap),
sub-pixel quantum efficiency variations,
charge transfer efficiency (CTE),
linearity range and saturation level.\\
Electronics: &
Bias, gain, ADC width, sampling (of charges), thermal noise.\\
Instrument: &
Dust in optics and optic distortions,
optical scale and detector pixel size (and their ratio to the typical
seeing -- spatial sampling),
file format (e.g. FITS \footnotemark),
and information beyond raw image (header keywords).\\
Environment: &
Signals originating from particle events (cosmic rays),
varying meteorological observing conditions (seeing), and
other atmospheric effects like sky illumination (moon) and extinction.\\
\hline
\end{tabular}
}
\caption{\label{t.ccd_properties}
Properties of raw CCD images and their origin.}
\end{table*}

We have expanded OIS with a standard reduction pipeline, a finding
algorithm for variable sources, a PSF-fitting photometry, and per
pixel error propagation for all steps of image processing.
We are able to give accurate errors for each photometric measurement
of a variable source beyond an estimate based on the noise in a single
difference image.
Our software has been developed to deal with time series obtained with
the 0.8\,m Wendelstein telescope and the Calar Alto 1.23\,m telescope
to monitor variable stars and find microlensing events towards the M\,31
bulge in the first instance.
However, we stress that many of the procedures presented here may also
be applied directly to other types of astronomical imaging.

\footnotetext{Flexible Image Transport System,
\citealp[see][]{1981A&AS...44..363W,1981SPIE..264..298G,1988A&AS...73..359G,1988A&AS...73..365H,1994A&AS..105...53P},
and NOST 100-2.0.}

The first part of this paper motivates the effort to implement per
pixel error propagation.
The second part gives a detailed description of our standard reduction
for CCD frames.
In the third part we present our image alignment procedure.
The fourth part describes the image convolution with OIS, the
detection procedure for variable sources and the relative photometry
of those sources.
In the last section we give the results of performance tests on
simulated images.

\section{Motivation for per pixel error propagation}
\subsection{Properties of raw CCD images}
\label{ss.ccd_properties}
We have to consider all properties of raw CCD images
(Tab.~\ref{t.ccd_properties})
before we can establish a reliable difference image analysis.
When estimating photometric errors all these effects and their errors
have to be considered again in addition to the photon noise induced by
the objects imaged.
To account for those additional photometric errors
the error has to be calculated and propagated for each pixel and every data
reduction step applied.
\subsection{Error propagation -- an analytic example}
\label{ss.intro_error}
We show that the errors resulting e.g. from flatfield calibration
(Sect.~\ref{ss.flatfield})
may be dominant for bright objects or inadequate flatfields which
cannot always be avoided.
Statistics for this effect on simulated images are shown in
Sect.~\ref{ss.flat_errors}.
(See Tab.~\ref{t.formulae} for common components of formulae used
throughout this article.)

\begin{table}
\centering
\tabcolsep0.5mm
\begin{tabular}{rcl}
\hline
\hline
$(x, y)$ & = & pixel coordinates,\\
$I(x, y)$ & = & value of pixel $(x, y)$ in image $I$,\\
$\delta_I(x, y)$ & = & corresponding absolute error,\\
$\tilde{I},\;\hat{I},\;\bar{I}$ & = &
sequence of indicators, that some reduction step\\
& & has been applied to $I$,\\
$n_I$ & = & integer number of e.g. images of type $I$,\\
$\sigma$ & = & the root mean square of a sample. \\
\hline
\end{tabular}
\caption{\label{t.formulae}Common components for the notation of
formulae.}
\end{table}

One can estimate this effect assuming
$n_I$ images, each with $I(x, y)$ detected photons per pixel, and
$\delta_I(x, y) = \sqrt{I(x, y)}$ error per pixel, so the preliminary
error estimate for a per pixel added stack
\begin{equation}
\tilde{I}(x, y) =
\sum_{j = 1}^{n_I}\frac{I_j(x, y)}{F(x, y)}
\approx
n_I \; \frac{I(x, y)}{F(x, y)}
\end{equation}
(where ``$\approx$'' is due to the fact that the $I_j$ will
vary, and $F(x, y) \approx 1$) is
\begin{equation}
\label{e.prelim_error}
\delta_{\tilde{I}}(x, y) \approx \delta_I(x, y) \; \sqrt{n_I} \;.
\end{equation}

Including as further assumptions a flatfield error $\delta_F(x, y)$,
a ratio of the relative errors between the flatfield and a single
image given by
\begin{equation}
\xi(x, y) = \frac{\delta_I(x, y)}{I(x, y)}
\left/ \frac{\delta_F(x, y)}{F(x, y)}\right. \;,
\end{equation}
and Gaussian error propagation,
we can calculate the impact of flatfield calibration
on the error $\delta_{\tilde{I}}(x, y)$ of the stack.
The propagated per pixel error depends on the individual
signal-to-noise ratios of the images and the flatfield, and on
dithering.
When stacking spatially undithered images, the flatfield error
$\delta_F(x, y)$ of the same pixel $(x, y)$ adds to each individual
image $I(x, y)$, so the flatfield error part of the stack is
not an independent error.
It is like stacking first and flatfielding subsequently:
\begin{equation}
\label{e.flatundithered}
\delta_{\tilde{I}}(x, y) \approx \delta_I(x, y) \;
\sqrt{n_I \left(1 + \frac{n_I}{\xi^2}\right)} \;.
\end{equation}
With spatially dithered (and digitally realigned) images we get
(by neglecting the effects of the alignment procedure)
\begin{equation}
\label{e.flatdithered}
\delta_{\tilde{I}}(x, y) \approx \delta_I(x, y) \;
\sqrt{n_I \left(1 + \frac{1}{\xi^2}\right)} \;,
\end{equation}
because different and therefore independent flatfield pixels add to
the per pixel error of the stack.

An example:
We consider an extended and bright object (e.g. the M\,31 bulge), and
twilight flatfields.
Here we get $1 < \xi < 3$ because of the difficulties in getting
twilight flats discussed in Sect.~\ref{ss.flatfield}.
In a dithered, $n_I = 5$ stack this yields an up to $41\%$ error
increment
and even a $145\%$ increment in an undithered, $n_I = 5$ stack,
both compared to the simple estimate of Eq.~\ref{e.prelim_error},
i.e. with respect to the Poissonian noise in the images alone.

Now we treat a complementary case:
When imaging faint objects in empty fields, the lack of proper
flatfields\footnote{It
is nearly impossible to get twilight flatfields for more than three
filters in one night; see Sect.~\ref{ss.flatfield}.}
pushes observers to build flatfields using the skylight in the
(dithered) images.
With $I(x, y) = \mathrm{object}(x, y) + \mathrm{sky}(x, y)$ and $n_F$
of $n_I$ images usable for building the flat
($n_F < n_I$ because of the objects and cosmics in the field)
we get
\begin{equation}
\delta_{\tilde{I}}(x, y) \approx \delta_I(x, y) \; \sqrt{n_I
\left[ 1 + \frac{1}{n_F} \left(1 +
\frac{\mathrm{object}(x, y)}{\mathrm{sky}(x, y)}\right)\right]}
\end{equation}
for the error in a stack, using
\begin{equation}
\frac{\delta_F(x, y)}{F(x, y)} = \frac{1}{\sqrt{n_F \mathrm{sky}}}
\quad\mbox{in Eq.~\ref{e.flatdithered}.}
\end{equation}
For $n_F = 5$ images, which can be used in every pixel for
flatfielding, and with an object which is much fainter than the sky,
this yields a $10\%$ error increment (here, and below again,
compared to the naive, Poissonian noise estimate of
Eq.~\ref{e.prelim_error}).
Assuming an object with flux comparable to the sky, the error increment
already exceeds $18\%$.

The per pixel propagation of errors gets even more important
when performing multi-pixel approximations,
as we do in some parts of our data reduction pipeline for the
difference image analysis.
When performing e.g. PSF-fitting one can enhance the accuracy by
including the appropriate error weights as stated in
Sect.~\ref{ss.interpolation_errors}.
\section{Standard reduction pipeline for images}
\label{s.standard}
\tabcolsep0.5mm
We start with the standard reduction for individual object frames.
All effects of Sect.~\ref{ss.ccd_properties}, which can be corrected
in single images, will be discussed.
We show a way how to compensate for these effects, and, in addition,
how these compensations affect the total error budget of each pixel.
\subsection{Saturation}
\label{ss.saturation}
Saturated pixels (due to ADC saturation or due to pixel charges at
full well capacity) have to be marked as wrong value pixels.
With a reasonable bias adjustment no pixel should have a zero value,
so we set saturated pixels to zero as a first step.
Since one cannot know if a saturated pixel has or has not
bloomed, the four directly adjacent neighbouring pixels should
also be treated like saturated.
Depending on the specific properties of a CCD, the
blooming assumption may be restricted to the two pixels
in-and-against the row-shift direction.
As we perform the bias correction (Sect.~\ref{ss.bias}) and the
initial calculation of an error frame (Sect.~\ref{ss.errorframe}) in
one step,
the zero marked pixels will still be recognised after the bias has
been subtracted.
Pixels with values beyond the linearity range of the CCD have to be
corrected, or, if not possible, also have to be treated like saturated
pixels.
Since we do not have to treat images with values in the non linear
regime we will not provide error propagation for this case.
\subsection{Bias correction}
\label{ss.bias}
The bias originates from the offset voltage of the CCD ADC.
Depending on its properties we remove it following two subsequent
approaches:
\begin{itemize}
\item
If the patterns in the bias frame are varying on short time scales,
or there is no pattern at all (just thermal noise),
we subtract only the $\kappa\sigma$-clipped mean
of a suitable part of the overscan
(i.e. a part of the overscan, which is identical in exposed frames and
dark, not exposed frames).
Since the bias is an additive constant and mostly a small
number the $\kappa\sigma$-clipped mean
($\kappa \approx 6$, to get rid of cosmics)
will be more accurate than the median.
\item
If the bias patterns are reproducible, we subtract in addition a
$\kappa\sigma$-median-clipped mean image of overscan corrected bias
frames.
(We check the reproducibility by looking for patterns in such a
mean image;
see Sect.~\ref{ss.flatfield} for details on
$\kappa\sigma$-median-clipping.)
\end{itemize}
Three types of errors have to be considered when correcting for the
bias:
\begin{itemize}
\item
Statistical error:
The readout noise consists of the thermal noise of the amplifier, and,
if present, of the dark current.
\item
Systematic error:
Unreproducible bias patterns may result from bad CCD
electronics or insufficient electronic shielding.
(Strong radio immission may even penetrate an excellent shielding.)
\item
Numerical error:
Error resulting from the numerical determination of the bias level.
\end{itemize}
Additional errors (e.g. count dependent bias) may arise because of a
bad CCD or bad electronics,
but dealing with those is beyond our scope.
Fortunately we also had no significant dark current to deal with on any of
the CCDs we have used so far.
\subsection{Generation of the error frame}
\label{ss.errorframe}
Our error frames are similar to a data-quality mask as used in
many pipelines, but differ in that their values are not representative
flags but actually numerical errors, with the exception of saturated
and bad pixels which are represented by simple flags (-1, 0).

Each pixel $(x, y)$ in an image $I$ has an initial error $\delta_I(x, y)$
resulting from the photon noise, the bias noise (i.e. readout
noise) plus the error in determination of the bias level:
\begin{equation}
\label{eq.bias}
\delta_{I}(x, y) =
\sqrt{\frac{\mathrm{counts}_I(x, y) - \mathrm{bias}_I}{\mathrm{gain}_I} +
\sigma_{\mathrm{bias}_I}^2 +
\frac{\sigma_{\mathrm{bias}_I}^2}{n_{\mathrm{bias}_I}} } \;,
\end{equation}
where\\
\begin{tabular}{lll}
$\mathrm{counts}_I$ & = &
flux of pixel $(x, y)$ in image $I$ in ADU,\\
$\mathrm{bias}_I$ & = & bias of the image,\\
$\mathrm{gain}_I$ & = &
$\frac{\mathrm{photons}_I}{\mathrm{ADU}}$ = conversion factor,\\
$\sigma_{\mathrm{bias}_I}$ & = &
we use the $\kappa(=6) \sigma$-clipped RMS of a\\
& & suitable part of the overscan as an estimation\\
& & for the bias noise (i.e. readout noise),\\
$n_{\mathrm{bias}_I}$ & = & number of pixels actually used for the
bias\\
& & determination.\\
\end{tabular}
\\
If using a small clipping factor ($\kappa < 3$) for the determination
of the bias noise,
$\sigma_{\mathrm{bias}_I}$ will be underestimated
and therefore has to be corrected by a factor
$1 / C_C$ where
\begin{equation}
\label{e.erf}
C_C = \mathrm{erf}(\kappa/\sqrt{2}) =
\frac{1}{\sqrt{2 \pi}} \int\limits_{-\kappa}^{\kappa}
e^{-\frac{1}{2}\hat\kappa^2} d\hat\kappa \;.
\end{equation}
Reproducible bias patterns can be determined with an accuracy only
limited by the applied numerical precision, so their error may be
neglected.\\
We mark saturated pixels in the error frame by setting them to minus
one,
which will be dominant in any error propagation from now on to prevent
the use of saturated pixels.
\subsection{Defective pixels}
\label{ss.defectpixel}
We identify defective pixels by checking the specific CCD documentation
as well as closely examining flatfields (for cold pixels, traps, and
coating defects) and darks (for hot pixels) with a large spread of
exposure times and counts if available.
All sorts of defective pixels we set to zero and approximate them later
(Sect.~\ref{ss.interpolate});
the error is also set to zero.
\subsection{Photosensitive region}
Since the overscan parts of a frame are not needed any more,
the frame is trimmed to its exposed part.
We also truncate any bad, first or last rows or columns.
On some CCDs the first row is broken and the edges of the
photosensitive area behave differently from the rest:
The effective size of the edge pixels can be up to $20\%$ larger than
the size of an average pixel.
\subsection{Low counts pixels}
\label{ss.locountpixel}
To ensure that only pixel values above an ignorance level will be
used, we set low count pixels and their error to zero.
This also accounts for the non linear range of a CCD due to a bad {\em
charge transfer efficiency} (CTE) at low pixel charges\footnotemark.
Like the defective pixels (Sect.~\ref{ss.defectpixel}) those pixels
will be estimated later (Sect.~\ref{ss.interpolate}), where it is
possible.
\footnotetext{Since we do not have to deal with short exposures of
bright objects in empty fields this procedure does not result in
ignoring the vast majority of pixels in a dataset.
When dealing with empty, low count backgrounds we propose following
e.g. \citet{1997wiley.book.....M} and adding a
{\em ``fat'' zero} by preflashing the CCD.}
\subsection{Flatfield calibration}
\label{ss.flatfield}
\subsubsection{Flatfielding basics}
In order to normalise the apparent photon sensitivity of all pixels in
a single CCD frame, a calibration image (flatfield image) has to be
built.
In an ideal case this would be the image of an extended, homogeneous,
flat, and white object at infinity.
The apparent photon sensitivity results from geometric size, coating,
and electronic properties of each single pixel, and, in addition, from
the inherent properties of the optics, and finally dust in the
optics.
Since there is no feasible way to get near the ideal case with
dome flats
(any additional optics which would project the nearby dome to infinity
would again add features to the calibration image),
we decided
to stretch our efforts to improve sky flats.
The daylight sky is the ideal case for flatfield images, but is much
to bright for broadband filter images.
Therefore the suitable time for getting sky flats is restricted to
dusk and dawn.
The superiority of twilight flatfields over dome flatfields is well
known and e.g. discussed in \citet{1991cacu.book.....B}, and
\citet{1986ARA&A..24..255M}.
Nevertheless,
dome flats and twilight flats can be used effectively in
combination as detailed in Sect.~\ref{sss.flatfieldcalculation}.
\subsubsection{Observation strategy for flatfields}
In order to minimise interfering effects like stars and photon noise,
we need at least five flatfield images per filter used,
fulfilling
these additional constraints:
\begin{itemize}
\item
The individual flatfield images should have count levels as high or
higher than the average of the observed object in a single frame,
as their noise adds to the data
(see Sect.~\ref{ss.intro_error} and \ref{ss.flat_errors}).
Nevertheless, flatfields with less counts are preferred to having less
than five flatfields.
\item
The exposures have to be long enough to avoid residuals of the
shutter (shutter pattern\footnotemark),
but still short enough to avoid saturation.
If the shutter movements show a predictable time dependency,
the flatfields can be deconvolved from the two-dimensional shutter
function in a simple way as proposed by \citet{1993A&A...278..654S}.
Unfortunately this is not the case with our observations for the
WeCAPP project \citep{2001A&A...0000..00R}.
Since the building of suitable flatfields already is the most (human)
time consuming part of the standard reduction phase,
we do not go into building shutter patterns for each individual
flatfield,
but simply exclude underexposed flats and try to manage with the
remaining.
\footnotetext{Assuming
an average flatfield charge per pixel of 160\,000 electrons
the shutter pattern will exceed one $\sigma$ photon noise,
if the exposure time is longer than 400 times the shutter movement
time (opening plus closing) of a non photometric shutter.
E.g. for an iris type shutter and a total shutter movement of 10\,ms
the exposure time has to exceed 4\,s just to have the additional
shutter error not bigger than the photon noise.
Since the shutter movement is a systematic effect, the combination of
many short time flatfields will even enhance that error by
increasing their fraction of all used flats and therefore amplifying
their impact on the resulting flatfield.}
\item
Flatfields should be taken of a blank field to have less
stars to be removed.
\item
The telescope focus should be well adjusted to minimise the number of
pixels affected by stars.
This is difficult for dusk flatfields, because one cannot make a
focus series of a star before starting with the flat series for
obvious reasons.
\item
All flatfields of one series should have some small
offsets in orientation,
so the $\kappa\sigma$-clipped median of the series will be clean of
stars.
We found an offset of $30''$ to be sufficient\footnotemark.
\footnotetext{Our blank fields are clean of bright stars within
the field of view of the cameras used (always less than $17 \times 17$
arcmin$^2$).
Since we also use only small, 1m-class telescopes there is no
considerable light contamination beyond the $30''$ limit of moderately
bright stars or galaxies at average flatfield exposure times.}
\item
Diffuse light contamination (i.e. reflections from inside the dome)
has to be avoided.
Neglecting this effect results in a mixture of dome
and sky flat with improper illumination\footnote{One
can check the impact of this light pollution effect by turning
on a weak dome light while taking an image in a new moon night.
We have improved considerably the situation at the Wendelstein
telescope by painting the interior dome surface black.}.
\end{itemize}
As one can easily estimate,
using the \citet{1993AJ....105.1206T} twilight formula,
it is impossible to get five flatfield images
(following above constraints)
for more than one filter,
if the time needed for CCD wipe and readout exceeds three minutes.
For those cases we found the restriction to only take
flat series for one filter per twilight, and to alternate the filters
for each twilight period,
to produce better results than the combination of multiple
insufficient flatfield series.
\subsubsection{Flatfield calculation and calibration}
\label{sss.flatfieldcalculation}
The final flatfield frame is built by combining individual flats
following this standard recipe:
\begin{enumerate}
\item
We follow the steps described in Sect.~\ref{ss.saturation} to
\ref{ss.locountpixel} for the individual flatfield images to build
appropriate flats $I$ and their correspondent error image $\delta_I$.
\item
\label{enum.flat.norm}
Then we normalise all flats by dividing them with the median $C_I$ of
a central area of the CCD
\begin{equation}
F = I / C_I \;.
\end{equation}
We determine the median only of the centre quarter of the full frame
in order to minimise the effect of differential illumination gradients
when combining the flats\footnotemark.
\footnotetext{The gradient due to vignetting within this central
area is median level $\pm$ 3\% in our images.
A higher gradient may still be acceptable as long as it is guaranteed
that the ``normalisation median'' lies within a well populated region
of the normalisation region's distribution.}
\item
\label{enum.flat.median}
To get rid of stars, cosmics, and differential illumination we build
a $\kappa\sigma$-clipped mean of the normalised flat frames $F$
\begin{equation}
\tilde{F}(x, y)=\frac{\sum\limits^{n_\mathrm{used}}_{F}
F(x, y)}{n_\mathrm{used}} \;.
\end{equation}
For each pixel the outliers are excluded by
$\kappa\sigma$-clipping,
but we use the median instead of the mean as reference for
the selection procedure because the median is less sensible to
occasional outliers;
the mean of the remaining pixels yields the final calibration factor.
We got the best results (least residuals) with a $\kappa = 1.0$,
which will preserve no more than two flats for most pixels.
With just five flatfields for combination a bigger $\kappa$ left
residuals at the $3\%$ level.
\item
\label{enum.flat.controlimage}
We control the result of step~\ref{enum.flat.median} by inspecting
all control frames $F_{\mathrm{control}} = F/\tilde{F}$.
They should neither show any signal $< 1$, like holes or shadows
around stars\footnotemark, nor illumination gradients.
Both effects would indicate residuals still left in the median
flatfield.
\footnotetext{Since the pollution of a flatfield image with stars
or cosmics is always an additional signal, the origin of residuals in
the control frames can be identified:
Residuals with a signal $> 1$ are correctly removed features of an
individual flat.
Residuals with a signal $< 1$, which also should be perceivable in
multiple control frames, must originate from the median flatfield.}
\item
Now we exclude those flats which cause residuals, and repeat
from step~\ref{enum.flat.median},
until there are no residuals left in the control images.
If this would lead into having too few flats, we add the median
flats of the days before and/or after to the median procedure.
\item
Since the pixel-to-pixel sensitivity variations are often exclusively
due to the CCD itself and therefore only a weak function of time, 
one can enhance a flatfield (i.e. improve its signal-to-noise ratio)
by combining the high spatial frequency,
pixel-to-pixel variation of several twilight periods
(or of very high signal-to-noise dome flats)
with the smoothed flatfield for the observed night,
whose low spatial frequency information can change from night to
night due to dust shadows etc.
\begin{enumerate}
\item
We smooth each median flat pixel within a box smaller than the typical
projected size of the short-time-scale variation due to dust:
Each pixel $(x_0, y_0)$ is smoothed according to
\begin{equation}
\tilde{F}_s(x_0, y_0) =
\frac{\sum\limits^{\mathrm{box}}_{x, y} \tilde{F}(x, y)}
{n_{\mathrm{box}}}
\; \mbox{, where}
\end{equation}
$n_{\mathrm{box}}$ = number of pixels in the smooth box.
\item
The pure pixel-to-pixel variation is given by
\begin{equation}
\tilde{F}_p(x, y) = \tilde{F}(x, y) / \tilde{F}_s(x, y) \;.
\end{equation}
\item
The pixel-to-pixel variation flats are now combined by building a
$\kappa\sigma$-clipped, median referenced mean as shown in
step~\ref{enum.flat.median}:
\begin{equation}
\hat{F}_p(x, y) = \frac{\sum\limits^{n_\mathrm{used}}_{\tilde{F}_p}
\tilde{F}_p(x, y)}{n_\mathrm{used}} \;.
\end{equation}
\item
The enhanced flatfield for the night can now be calculated via
$\hat{F}(x, y) = \tilde{F}_s(x, y) \; \hat{F}_p(x, y)$.
\end{enumerate}
This procedure will only be applied, if the observational
circumstances will lead to a gain in the signal-to-noise ratio of the
resulting flatfield.
A comparable technique has been used to build high quality HST
flatfields \citep{1994AJ....108.2362R}.
\item
It is possible to check the gain (after a change of detector or an
observational gap) using following correlation:
\begin{equation}
\mathrm{gain} = \frac{1}{C_I \; \sigma_{F/\tilde{F}}^2}\;
\left(= \frac{C_I}{\sigma_{I/\tilde{F}}^2}\right)
\;\mbox{, where}
\end{equation}
\begin{tabular}{lll}
$F, C_I, \tilde{F}$ & = & see steps~\ref{enum.flat.norm} and
\ref{enum.flat.median},\\ 
$\sigma_{F/\tilde{F}}$ & = &
RMS of all pixels of a control image $F/\tilde{F}$,\\
$\sigma_{I/\tilde{F}}$ & = &
RMS of an alternative control image $I/\tilde{F}$.\\
\end{tabular}\\
This is only an approximation neglecting the readout noise and
assuming identical pixel sensitivities.
\\
\item
All data frames are divided by the finally accepted flatfield:
$\tilde{I}(x, y) = I(x, y) / \hat{F}(x, y)$.
\end{enumerate}
We are still considering options to build a full automated flatfield
evaluation procedure analysing the control images.

The whole flatfielding procedure can also be performed by
standard astronomical data reduction software but without error
propagation.
\subsubsection{Error propagation}
\label{sss.flatfieldpropagation}
The statistical error is propagated as follows;
all approximations are only given to illustrate the impact of the
respective reduction step and assume a normalised flatfield flux
$\approx 1$ and negligible pixel-to-pixel and image-to-image variation
of the error:
\begin{enumerate}
\item
Normalisation, error of pixel $(x, y)$ in normalised flat $F$:
\begin{eqnarray}
\delta_F(x, y) & = &
F(x, y) \; \sqrt{\left(\frac{\delta_I(x, y)}{I(x, y)}\right)^2 +
\left(\frac{\delta_{C_I}}{C_I}\right)^2}
\; ,
\end{eqnarray}
where\\
\begin{tabular}{lll}
$F(x, y)$ & = & flux of pixel $(x, y)$ in normalised flat $F$,\\
$I(x, y)$ & = & flux of pixel $(x, y)$ in bias corrected flat $I$,\\
$\delta_I(x, y)$ & = & error of pixel $(x, y)$ in bias corrected flat $I$\\
& & defined in Eq.~\ref{eq.bias},\\
$C_I$ & = & normalisation factor, see
Sect.~\ref{sss.flatfieldcalculation}, step~\ref{enum.flat.norm},\\
$\delta_{C_I}$ & = & error of normalisation factor, we use a\\
& & standard error of the median\\
& = & $\frac{\sigma_{\mathrm{med}}}{\sqrt{n_\mathrm{med}}} =
\sqrt{\frac{\sum\limits^{n_\mathrm{med}}_{x, y}
\left(C_I - I(x, y)\right)^2}
{(C_I n_\mathrm{med})^2}}$ ,\\
$n_{\mathrm{med}}$ & = & number of pixels used to build the median.\\
\end{tabular}
\\
An uniform CCD, homogeneously illuminated flatfields, and a suitable
(large) tailored area for the determination of $C_I$ altogether will
lead to a negligible $\delta_{C_I}$, which can be seen via
\begin{eqnarray}
\delta_F(x, y) & \approx &
\frac{\delta_I(x, y)}{C_I} \;
\sqrt{1 + \frac{1}{n_{\mathrm{med}}}}
\; \mbox{, assuming}\nonumber\\
\delta_{C_I} & \approx & \frac{\delta_I(x, y)}{C_I \;
\sqrt{n_{\mathrm{med}}}} \quad \mbox{and} \quad
C_I \approx I(x, y)\;.\nonumber
\end{eqnarray}
If using $\kappa\sigma$-clipping $\sigma_{\mathrm{med}}$ has to be
corrected as shown in Sect.~\ref{ss.errorframe} using
Eq.~\ref{e.erf}.
\item
\label{enum.median.flaterror}
Building of median flatfield,
error of pixel $(x, y)$ in median flat $\tilde{F}$:
\begin{eqnarray}
\label{e.medflaterror}
\delta_{\tilde{F}}(x, y) & = &
\frac{\sqrt{\sum\limits^{n_\mathrm{used}}_{F} \delta_F^2(x, y)}}
{\mathrm{erf}\left(\frac{\kappa}{\sqrt{2}}\right) \;
\sqrt{n_\mathrm{used} \; n_\mathrm{total}}}\\
& \stackrel{\kappa=1}{\approx} &
\delta_F \; \sqrt{\frac{2}{n_\mathrm{total}}}
\stackrel{\kappa>3}{\approx} \frac{\delta_F}{\sqrt{n_\mathrm{total}}}
\;\mbox{, where}\nonumber
\end{eqnarray}
\begin{tabular}{lll}
$n_\mathrm{used}$ & = & remaining number of flats after clipping\\
& & used for a specific median clipped mean\\
& & pixel,\\
$n_\mathrm{total}$ & = & total number of flats used for clipping,\\
$\mathrm{erf}\left(\frac{\kappa}{\sqrt{2}}\right)$ & = &
see Eq.~\ref{e.erf}, and\\
$\kappa$ & = & clipping factor.
\end{tabular}
\\
Ignoring the effect of using preselected,
$\kappa\sigma$-clipped pixels to calculate the mean calibration pixel
would lead to a significant misestimation of the resulting error for
small $\kappa$.
The assumption of normally distributed values is crude but still fair
(see Sect.~\ref{ss.flat_errors}).
\item
The error of the flatfield enhancement (if applied):\\
\begin{enumerate}
\item
The error of a smoothed pixel $(x_0, y_0)$ built by averaging
independent pixels is
\begin{equation}
\delta_{\tilde{F}_s}(x_0, y_0) =
\frac{\sqrt{
\sum\limits^{\mathrm{box}}_{x, y} \delta^2_{\tilde{F}}(x, y)}}
{n_{\mathrm{box}}}
\approx
\frac{\delta_{\tilde{F}}}{\sqrt{n_{\mathrm{box}}}} \;.
\end{equation}
\item
The error for the pixel-to-pixel flatfield is given by
\begin{eqnarray}
\delta_{\tilde{F}_p}(x, y) & = &
\tilde{F}_p(x, y) \; \nonumber\\
& & \times \sqrt{
\left(\frac{\delta_{\tilde{F}}(x, y)}{\tilde{F}(x, y)}\right)^2 +
\left(\frac{\delta_{\tilde{F}_s}(x, y)}{\tilde{F}_s(x, y)}\right)^2}\\
& \approx & \delta_{\tilde{F}} \;
\sqrt{1 + \frac{1}{n_{\mathrm{box}}}} \;. \nonumber
\end{eqnarray}
\item
Combining the pixel-to-pixel flats yields
(as in step~\ref{enum.median.flaterror})
\begin{eqnarray}
\delta_{\hat{F}_p}(x, y) & = &
\frac{\sqrt{\sum\limits^{n_\mathrm{used}}_{\tilde{F}_p}
\delta^2_{\tilde{F}_p}(x, y)}}
{\mathrm{erf}\left(\frac{\kappa}{\sqrt{2}}\right) \;
\sqrt{n_\mathrm{used} \; n_\mathrm{total}}}\\
& \stackrel{\kappa=1}{\approx} &
\delta_{\tilde{F}_p} \; \sqrt{\frac{2}{n_\mathrm{total}}}
\stackrel{\kappa>3}{\approx} \frac{\delta_{\tilde{F}_p}}
{\sqrt{n_\mathrm{total}}} \;. \nonumber
\end{eqnarray}
\item
The error of the enhanced flatfield can be calculated with
\begin{eqnarray}
\delta_{\hat{F}}(x, y) & = &
\hat{F}(x, y) \; \\
& & \times \sqrt{
\left(\frac{\delta_{\tilde{F}_s}(x, y)}{\tilde{F}_s(x, y)}\right)^2 +
\left(\frac{\delta_{\hat{F}_p}(x, y)}{\hat{F}_p(x, y)}\right)^2}
\nonumber\\
& \stackrel{\kappa>3}{\approx} & \delta_{\tilde{F}} \;
\sqrt{\frac{1}{n_{\mathrm{box}}} +
\frac{1 + 1 / n_{\mathrm{box}}}{n_{\mathrm{total}}}} \;. \nonumber
\end{eqnarray}
\end{enumerate}
\item
Flatfield division, error of pixel $(x, y)$ in flatfield calibrated
image $\tilde{I}$:
\begin{eqnarray}
\delta_{\tilde{I}}(x, y) & = &
\tilde{I}(x, y) \; \nonumber\\
& & \times \sqrt{
\left(\frac{\delta_I(x, y)}{I(x, y)}\right)^2 +
\left(\frac{\delta_{\hat{F}}(x, y)}{\hat{F}(x, y)}\right)^2}\\
& \approx &
\sqrt{\delta^2_I + (\delta_{\hat{F}} \; I)^2} \;. \nonumber
\end{eqnarray}
\end{enumerate}
Minor systematic errors are neglected here:
\begin{itemize}
\item
The flatfield response of a CCD is a strong function of colour.
This results in a systematic error when calibrating stars with colours
different from the sky on an image,
and gets worst when the compared objects have highly different
colours.
\item
The geometric distortion introduced with the variation of CCD pixel
size is ignored in our flatfielding procedure.
A position estimate will have a systematic error according to this,
if determining positions of objects in undersampled images
(e.g. due to extraordinary good observing conditions and therefore
very sharp PSFs).
\item
Since we ignore the individual geometric sizes of CCD
pixels the integrated photometry of a flatfielded image may be
corrupted
(not exceeding 0.5\% in a single pixel of our images).
\end{itemize}
All those errors could be compensated, but will have at most
a minor (but detected) influence on our data because of OIS,
relative profile fitting photometry,
dithered image stacks and error propagation
(Sect.~\ref{ss.convolution}, \ref{ss.kernel}, and \ref{ss.photometry}).
\subsection{Filtering of cosmic rays}
\label{ss.cosmicfilter}
There are two major reasons, why we have to correct for the image
contamination by particle events (so called {\em cosmics\/}):
We use small, 1m-class telescopes for our observational projects
\citep[e.g.][]{2001A&A...0000..00R}
and therefore have integration times of half an hour or longer.
We want to find variable sources and establish light curves for every
pixel of an observed field.
So we have to identify individual cosmics in single frames
automatically at least, or better clean the images of cosmics and
account for the error this procedure might introduce in a stack of
images.
\subsubsection{Common and literature filters}
Existing filtering techniques for particle events either rely on
median stacking of multiple well aligned images
\citep[i.e. see][]{1994PASP..106..798W}
or compare each pixel value with the median of its
neighbours and define pixels with a sharpness ratio above a
deliberately set value as cosmic.
The first approach does not work at all if there are no multiple
images available or the sample of images to be stacked has different
observational features (variable sky, extinction or
PSF).
Aligning images will always spread and diffuse cosmics and therefore
obscure them.
The latter technique gets into trouble with noisy images, undersampled
images and multiple-pixel cosmics for obvious reasons.

Trainable cosmic classifiers i.e. as described in
\citet{1995PASP..107..279S} have the advantage of also being
applicable to undersampled data 
but rely on subjectively defined training sets
which are difficult to create for a large spread of different
telescope, camera, and detector configurations in addition to the
great range of observing conditions.
An interesting new approach is presented in
\citet{2000PASP..112..703R}.
Since this method relies on an accurate PSF and sky determination and
the author does not refer to possible problems in heavily crowded fields
we did not use it.
Furthermore, this technique is in principle only sensitive to single
pixel events,
which we found to be not the common case.
In fact most cosmics seem to have a major-to-minor axis
ratio\footnote{We
have checked this with the control output of our filter code.
It gives the major and minor axis full width half maximum of the
Gaussian fit function for every cosmic replaced.}
greater than two.
Multiple-pixel events, which can be filtered with our technique (see
below) in one pass,
have to be detected iteratively and with decreasing efficiency 
with the \citeauthor{2000PASP..112..703R} technique.
\subsubsection{Gaussian filter}
We apply a straightforward Gaussian filter to every single
image:
We fit five-parameter Gaussians to all local maxima of an image.
If the width along one axis of the fitting function is smaller than a
threshold
(which has to be chosen according to the PSF)
and, in addition,
the amplitude of the fitting function exceeds the expected noise by a
factor
(which has to be chosen according to the additional noise due to
crowding, see Sect.~\ref{ss.cosmic_errors} for details),
we replace the pixels with the fitted surface constant,
where the fitting function exceeds this constant by more
than two times the expected photon noise.
In the following we describe the algorithm in detail:
\begin{enumerate}
\item
Because of code speed improvements which rely on some symmetries in
the fitting function the tested cosmic candidate has to be in the
centre of a $7 \times 7$ pixels array.
In order to be applicable also on the first and last three rows and
columns we add a border surrounding the exposed frame filled with
zero value pixels.
Since we do not want to lose too much of the images when
shifting them later (Sect.~\ref{s.align}) we chose to enlarge the
images not only with a three pixels but with a
20 pixels border instead.
\item
\label{enum.cosmic.max}
Now, first we search for all local maxima $(x_0, y_0)$ in the image,
but ignore those with either a large error\footnote{We
found $\gamma \approx 3$ to be an empirically suitable factor.}
$\left(\delta^2(x_0, y_0) > \gamma^2 \; \frac{\mathrm{signal(x_0,
y_0)}}{\mathrm{gain}}\right)$
or more than two saturated neighbours\footnote{Cosmics
candidates close to saturated pixels need a special treatment, see
step~\ref{enum.cosmic.saturate}.
Because we always consider the possibility of blooming
(see Sect.~\ref{ss.saturation}),
only for pixels with at least three saturated
marked pixels (of the eight possible neighbours)
one pixel (of the four directly adjacent) is really saturated.}
or with less than four (of eight possible) valid
neighbours\footnote{The
Gaussian fit of step~\ref{enum.cosmic.fit} gets unstable
when fewer than half
of the pixels adjacent can be used in the fitting algorithm.}.
\item
\label{enum.cosmic.fit}
Then we perform a propagated error weighted, least-squares fit, assuming
a five-parameter, two-dimensional Gaussian fitting function
centred on these local maxima ($x_0, y_0$) coordinates in $7 \times 7$
pixels subarrays:
We determine a surface constant $C$, an amplitude $A$, a rotation
angle $\alpha$, a major and a minor axis full width half maximum
($x_\mathrm{fwhm}$ and $y_\mathrm{fwhm}$) of the fitting function
$f_{\mathrm{gauss}}$
giving the flux of a pixel $(x, y)$
\begin{eqnarray}
f_\mathrm{gauss}(x, y) & = & C +\\
& &
A \exp
\left[ - 4 \ln 2 \left( \frac{x'^2 }{x_\mathrm{fwhm}^2} +
\frac{y'^2}{y_\mathrm{fwhm}^2}\right)\right], \nonumber
\end{eqnarray}
where
\begin{eqnarray*}
x'(x, y) & = & \left(x - x_0\right) \cos \alpha + \left(y - y_0\right)
\sin \alpha,\\
y'(x, y) & = & \left(y - y_0\right) \cos \alpha - \left(x - x_0\right)
\sin \alpha.
\end{eqnarray*}
\item
All Gaussians with an amplitude of $t_\mathrm{limit}$ times the
propagated error of the centre pixel and a full width half maximum in
one axis smaller than a limiting $\mathrm{FWHM}_{\mathrm{cosmic}}$ are defined as
\em cosmic\rm \/:\\
$\left(A > t_\mathrm{limit} \; \delta(x_0,
y_0)\right) \wedge$\\
$\left[\left(x_\mathrm{fwhm} <
\mathrm{FWHM}_{\mathrm{cosmic}}\right) \vee
\left(y_\mathrm{fwhm} < \mathrm{FWHM}_{\mathrm{cosmic}}\right)\right]$
\item
We have to perform a sanity check on the fitting function:
The surface constant $C$ and the amplitude $A$ must be positive;
$\bar{\chi}^2$ of the fit must be close to unity\footnote{This
corresponds to a compatible fitting function and correct error
weights.}:
\begin{equation}
C > 0 \;\wedge\; A > 0 \;\wedge\;
\bar{\chi}^2_\mathrm{fit} \simeq 1
\mbox{, where}
\end{equation}
\begin{eqnarray}
\label{e.chisquare}
\bar{\chi}^2_\mathrm{fit} & = &
\mbox{the reduced $\chi^2$ of the fit}\nonumber\\
& = & \frac{\sum\limits^{\mathrm{fitbox}}_{x, y}
\left(\frac{f(x, y) -
\tilde{I}(x, y)}
{\delta_{\tilde{I}}^2(x, y)}\right)^2}
{n_\mathrm{dof}} \;,\\
n_\mathrm{dof} & = &
\mbox{degrees of freedom for the fit,}\nonumber\\
 & = & (7 \times 7) - 5 \;\mbox{ here.}
\end{eqnarray}
\item
\label{enum.cosmic.substitute}
Now we substitute pixels $(x, y)$ where the Gaussian
fitting function $f_\mathrm{gauss}$ gives a value larger than the
assumed signal plus two times the assumed photon noise\\
$f_{gauss}(x, y) > C + 2 \sqrt{\frac{C}{\mathrm{gain}}}$.
\item
\label{enum.cosmic.error}
The new error of the substituted pixels is set to
\begin{equation}
\sqrt{\delta^2_{\tilde{I}}(x, y) + \bar{\chi}^2_\mathrm{fit}
\; \frac{C}{\mathrm{gain}}}
\;\mbox{, where}
\end{equation}
\begin{tabular}{lll}
$\delta_{\tilde{I}}(x, y)$ & = & propagated old error of pixel,\\
$\bar{\chi}_\mathrm{fit}$ & = & accuracy of the Gaussian fitfunction\\
 & & ($\bar{\chi} \approx 1$ for a perfect fit).\\
\end{tabular}

This will be large enough to prevent an incautious use of the
replaced pixels in the following.
\item
We then repeat this procedure for the areas with cosmics found
beginning with step~\ref{enum.cosmic.max} until no more cosmics are
found.
\item
\label{enum.cosmic.saturate}
Finally we try to find and replace cosmics near saturated pixels with
a similar, just in some details more sophisticated technique:
The Gaussian fit starts centred on the saturated region but the
centre position is added to the list of free parameters.
We ignore the saturated region for the fitting and
the replacement procedure.
For overall stability reasons we have to use closer constraints for
the sanity check.
Since saturated pixels due to cosmics will not be treated at all,
because they are flagged as ``dominant bad pixels'',
step~\ref{enum.cosmic.saturate} might be readjusted, if dealing with
shallow-well CCDs;
saturated regions can be replaced, but then saturated objects may be
mistaken for a cosmic.
\end{enumerate}
We found that a $t_\mathrm{limit} = 8.0$ and a $\mathrm{FWHM}_{\mathrm{cosmic}}
= 1.5$ works fine in any well sampled image.
However, in some extraordinary good seeing images
(with an average stellar PSF FWHM $\leq 2.0$ pixels)
we had to specify a limiting $\mathrm{FWHM}_{\mathrm{cosmic}} = 1.3$ to avoid the
deletion of stars.
All fixed fitting and substitution constants were adjusted
in order to get an accurate and reliable filter for cosmic rays
for all our images.
The sensitivity parameters $t_\mathrm{limit}$ and
$\mathrm{FWHM}_{\mathrm{cosmic}}$ have nevertheless to be adjusted in general to
the observational and object properties to reach the best compromise
between false alarm and fail detection rate
(Sect.~\ref{ss.cosmic_errors}).
\subsection{Approximation of bad pixel areas}
\label{ss.interpolate}
Pixels with value and error set to zero will now be replaced.
We use a distance and error weighted linear approximation of the
closest neighbours.
The fitting box is selected as small as possible with the restriction
that more than $2/3$ of the fitting box pixels minus the centre pixel
must be valid pixels and the fit box may not be larger than an arbitrary
limit which we set according to the spatial resolution of a specific
imaging system.
If even the largest possible box does not apply to the first
criterion the pixel is considered as isolated
and not replaced at this point\footnotemark.
\footnotetext{Isolated pixels still can be replaced with the mean value
of corresponding pixels in the remaining images of a stack
(Sect.~\ref{s.align}).
We do not apply this method, which relies on a perfect photometric
alignment of the to-be-stacked frames before any image convolution has
been applied.
If we need a measurement in the lost area, we build a separate stack
without the image with isolated pixels in the interesting region.}
Each replaced pixel $(x_0, y_0)$ of an image $\tilde{I}$ gets an
error calculated from the individual pixel errors of the fitting
box
\begin{equation}
\delta_{\tilde{I}}(x_0, y_0) =
\bar{\chi}_\mathrm{fit} \;
\frac{\sum\limits^{n_\mathrm{used}}_{x, y}
{\delta_{\tilde{I}}(x, y)}}{n_\mathrm{used}}
\;\mbox{, where}
\end{equation}
\begin{eqnarray*}
n_\mathrm{used} & = &
\mbox{number of pixels used for fit, and}\\
\bar{\chi}_\mathrm{fit} & = &
\mbox{defined according to Eq.~\ref{e.chisquare}.}
\end{eqnarray*}
A larger fitting box has fewer close base points and
therefore raises the uncertainty of the linear approximated
substitute,
so we use an average error of the input parameters
times the quality of the fit ($\bar{\chi}_\mathrm{fit}$)
and not only the error of the calculated value
($\propto1/n_\mathrm{used}$).
Like in Sect.~\ref{ss.cosmicfilter}, step~\ref{enum.cosmic.error}
this prevents an incautious use of the replaced pixels in
the following.\\
\section{Position alignment and stacking}
\label{s.align}
Up to now our reduction pipeline can be applied to any image
regardless of its scientific application.
When shifting images in order to stack them the PSF and even the flux
may be altered.
It is important to conserve the PSF, because the image convolution
approach that we adopt performing differential photometry relies
crucially on it.

In any case the alignment of images follows a four step procedure:
\begin{enumerate}
\item
First we determine the coordinates of reference objects in every image,
\item
then we calculate the coordinate transformation to project an image
onto the reference frame,
\item
subsequently we project the images into the reference frame coordinate
grid, and
\item
finally we stack the images.
\end{enumerate}
\subsection{Position of reference stars by PSF-fitting}
\label{ss.refstars}
In order to obtain the coordinates of reference objects in all images
we perform interactively a seven-parameter Gaussian fit.
We begin with the reference frame:
About 20 stars with a high signal-to-noise ratio and well distributed
over the frame would be sufficient but we use 50,
because stars close to the frame border may be missing on some images.
We continue with selecting at least one reference object in every
image manually,
the rest will be found automatically\footnotemark.
\footnotetext{In case the imaging device provides an accurate World
Coordinate System (WCS) information all of the reference stars could
be found automatically.}
The lists with the reference objects in the reference frame and the
first reference object of the unshifted images are used to recognise
the reference objects in each image,
to determine their position, and finally to calculate the projection
parameters.
\subsection{Translation of image coordinates -- determination of
a linear projection}
\label{ss.projection}
With the telescopes and cameras used in our observing campaigns, we
found a linear relation to be sufficient.
We easily match 50 stars all over a $8'\:\mathrm{x}\;8'$ field within
$1/20\;''$ in the mean.
Since there was no significant optical field distortion,
it was not necessary to use a non linear relation.
We determine a $2 \times 2$ linear matrix and a
two-dimensional translation vector
\begin{equation}
\left(\begin{array}{c}x'\\y'\end{array}\right)=
\left(\begin{array}{cc}a_{11}&a_{12}\\a_{21}&a_{22}\end{array}\right)
\left(\begin{array}{c}x\\y\end{array}\right) + 
\left(\begin{array}{c}t_1\\t_2\end{array}\right)
\end{equation}
with a least-squares fit.
It matches the positions of reference stars in the reference
system with the positions in the unshifted image with this
six-parameter relation.
\subsection{Shifting of images}
The technique of Variable-Pixel Linear Reconstruction or {\em
drizzling}
\citep{1997SPIE.3164..120F,1997adass...6..147H,1997AAS...191.4107M}
offers the possibility to add images while preserving both the flux
and the PSF.
In undersampled images one might even enhance the resolution and
therefore gain signal to noise.
Unfortunately this technique requires some image properties to be
applicable and
these are very difficult to obtain with ground-based telescopes:
There must be no variations in sky, extinction and PSF, and
there should be a uniform spatial sampling in the sub-pixel pointing.
If those requirements are missed, the flux will still be preserved,
but the PSF may get very distorted.
So aperture photo\-metry might still work very well, but since we use
PSF convolution and a PSF-dependent photo\-metry, we had to think for
an alternative with less observational constraints.

Our shift algorithm preserves PSF in unstacked and sometimes
undersampled frames.
We found that a 16-parameter, $3^\mathrm{rd}$-order polynomial
interpolation with 16 pixel base points does apply to our needs.
A $2^\mathrm{nd}$-order polynomial still smoothes the images, whereas
a $4^\mathrm{th}$-order polynomial does no better PSF conservation
compared to the $3^\mathrm{rd}$-order polynomial.
Since the number of parameters of the polynomial is matched with the
input base points, no least-squares fit is needed;
the polynomial can be calculated analytically.

The flux interpolation for non-integer-value coordinates $(x, y)$
is calculated with a polynomial
\begin{equation}
\label{e.interpolynom}
p(x,y)=\sum_{i=0}^{3} \sum_{j=0}^{3} a_{ij} x^i y^j
\;\mbox{, where}
\end{equation}
\begin{tabular}{lcl}
$i, j$ & $=$ &
index for subscript, and exponent for superscript.
\end{tabular}
For a region with $4 \times 4$ pixels this yields 16 linear equations
\begin{equation}
p(x_k,y_k) =
\tilde{I}(x_k,y_k) \;\mbox{, where } 1 \leq k \leq 16,
\end{equation}
so the coefficients $a_{ij} = a_{ij}(\tilde{I}(x_k,y_k))$ can be
calculated by solving the matrix equation.
The error is calculated using Gaussian error propagation
\begin{eqnarray}
\delta_p(x,y) & = & \sqrt{\sum_{i=0}^{3} \sum_{j=0}^{3}
(x^i y^j)^2 \; \delta_{a_{ij}}^2}
\;\mbox{, where}\\
\delta_{a_{ij}} & = & \sqrt{
\sum_{k=1}^{16}\left(\frac{\partial a_{ij}(\tilde{I}(x_k,y_k))}
{\partial \tilde{I}(x_k,y_k)}\;
\delta_{\tilde{I}}(x_k,y_k)\right)^2}
\;.\nonumber
\end{eqnarray}
\subsection{Stacking images}
In order to stack images we simply sum up the aligned frames,
but exclude outliers
(satellite polluted, occasional bad seeing or bad focus images)
and if convenient split into groups
(i.e. if many observations available of one night or very variable
PSFs etc.).
When searching for variable sources one always has the choice of
trading time resolution for a deeper magnitude limit.
One could also get both, if looking for periodic variables,
but then one has to
test every possible period with different stacks and
one may also have to
stack images with very different observational properties.
Pixel defects still marked with zero in any frame
(like saturated or not interpolatible pixels)
are set to zero in a stack.

The error of a pixel $(x, y)$ in an image stack $S$ consisting of
$n_{\tilde{I}}$ images propagates as
\begin{equation}
\delta_{S}(x, y) = \sqrt{\sum\limits^{n_{\tilde{I}}}_{\tilde{I}}
\delta_{\tilde{I}}^2(x, y)} \;.
\end{equation}
\section{Detection of variable objects}
\tabcolsep0.3mm
In order to detect variable sources we are using DIA
({\em Difference Image Analysis}\/), a method proposed by
\citet{1996AJ....112.2872T}.
The idea of DIA is to subtract two positionally and photometrically
aligned frames which are identical except for variable sources.
The resulting difference image should then be a flat noise frame,
in which only the variable point sources are visible.

A crucial point of this technique, apart from position
registration (Sect.~\ref{s.align}) is the requirement of a perfect
matching of the PSFs between the two frames.
For this purpose we apply a method, called OIS
({\em Optimal Image Subtraction}\/) developed by
\citet{1998ApJ...503..325A}.
\subsection{Photometric Alignment}
\label{ss.photalign}
Although OIS implements the photometric alignment from the reference
frame to each other frame,
it is useful to align all frames photometrically before matching the
PSF.
This ensures that the whole light curves are photometrically
calibrated to a standard flux.

All frames $S(x, y)$ are connected to a flux standard frame
$S_0(x, y)$ by\footnote{To be consistent with the
\citet{1998ApJ...503..325A} notation we now call images again plain
$I$.}
\begin{equation}
S_0(x, y) \approx a \; S(x, y) + b = I(x, y) \;.
\end{equation}
The scaling factor for different exposure times and atmospheric
extinction $a$ and background sky light $b$
are determined in a simple and crude way.
As a first step we remove all obvious bright stars from our field and
replace them by a plane representing the surrounding background level.
By replacing in both frames each pixel with the median count rates of
$21 \times 21$ pixels subsections, the influence of a
different PSF between the frames on the determination of $a$ and $b$
is minimised.
With these replaced pixel images we calculate values for $a$ and $b$ by
solving the least-squares problem.

The error frame is calculated using Gaussian error propagation
\begin{equation}
\delta_I(x, y) = \sqrt{S^2(x, y) \; \delta_a^2 +
a^2 \; \delta^2_{S}(x, y) + \delta_b^2} \;.
\end{equation}
\subsection{Convolution with differential background subtraction}
\label{ss.convolution}
One advantage of OIS is the possibility to fit
differential background variations simultaneously with the PSF 
between the frames.

Including a background term the convolution
equation, which transforms the PSF with smaller FWHM of the
reference frame $R$ to the PSF of an image $I$, is of the form 
\begin{equation}
I(x, y) \approx R(u, v) \otimes K(u, v) + bg(x, y) = \tilde{R}(x, y)
\;,
\end{equation}
where $(R\otimes K) (x,y)= \sum\limits_{u, v} R(x+u,y+v) K(u,v)$ .\\
The convolution kernel $K(u,v)$ and the background term $bg(x,y)$ are 
decomposed into basis functions
\begin{eqnarray}
K(u,v) & = & \sum_{i=1}^n a_i B_i(u,v)
\;\mbox{, and}\nonumber\\
bg(x,y) & = & \sum_{i=n+1}^{n+n_{bg}} a_i x^{p_i} y^{q_i}
\; ,\nonumber
\end{eqnarray}
where $n$ is the total number of coefficients of $K(u,v)$ and $n_{bg}$
is the corresponding number for the background term $bg(x,y)$.
The exponents $p_i$ and $q_i$ are fixed integers.

$K(u,v)$ is determined by solving the least-squares problem:
\begin{eqnarray}
\chi^2 & = &
\sum_{x,y} \frac{1}{\sigma_{x,y}^2}
\left[(R\otimes K) (x,y) + bg(x,y) - I(x,y) \right]^2 \\
& \stackrel{!}{=} & \min \nonumber
.
\end{eqnarray}
By setting $\frac{\partial\chi^2}{\partial a_j}\stackrel{!}{=}0$
these equations transform into
\begin{eqnarray}
&&\sum_i a_{i} \sum_{x,y}
\frac{1}{\sigma_{x,y}^2} C_i(x,y) C_j(x,y) \\
&&\quad = \sum_{x,y} \frac{1}{\sigma_{x,y}^2} I(x,y) C_j(x,y)
\; \mbox{, where}\nonumber
\end{eqnarray}
\begin{equation}
C_i(x,y) = \left\{
\begin{array}{l@{\quad}l}
R(u,v) \otimes B_i(u,v) & i=1,\dots,n \\
x^{p_i} y^{q_i} & i=n+1,\dots, \\
& \quad \quad n+n_{bg}
\end{array} 
\right.
\end{equation}
The problem is reduced to the solution of the following matrix
equation for the $a_i$ coefficients
\begin{equation}
  \sum_{i} a_i \; M_{ij} = V_j
\quad ,\qquad
  \underline{\underline{M}} \; \underline{a} = \underline{V}
\end{equation}
where the matrix elements are defined according to
\begin{eqnarray} 
M_{ij} & = & \sum_{x,y} \frac{1}{\sigma_{x,y}^{2}}
C_{i}(x,y) \; C_{j}(x,y) \;,\\
V_{j} & = & \sum_{x,y} \frac{1}{\sigma_{x,y}^{2}} I(x,y)
\; C_j(x,y) \;.
\end{eqnarray}
\subsection{Application to data}
\label{ss.kernel}
We adopt Gaussians modified with polynomials of order $p$
as a suitable kernel model
as proposed by \citet{1998ApJ...503..325A}
\arraycolsep0.4mm
\begin{eqnarray*}
K(u,v) & = & \sum_i a_i B_i(u,v) = \\
& = &
\sum\limits_{l}e^{-\frac{u^2+v^2}{2\sigma_l^2}}
\sum\limits_{j=0}^{p_l}\sum\limits_{k=0}^{p_l-j} a_{ljk}\;u^j v^k \; .
\end{eqnarray*} 
In order to get an optimal kernel, which can be of a complicated
shape, we use the combination of three Gaussians
\citep{1998ApJ...503..325A},
each with a different width $\sigma$.
This leads to the following $49$ parameter decomposition of the
polynomials of $K(u,v)$:
\begin{equation}
\begin{array}{lll}
\sigma_1=1,& p_1=6: &
(a_1 + \dots + a_{7} v^6 + \dots + a_{28} u^6)\\
\sigma_2=3,& p_2=4: &
(a_{29}+ \dots +a_{33} v^4+ \dots +a_{43} u^4)\\
\sigma_3=9,& p_3=2: &
(a_{44} + \dots + a_{46} v^2 + \dots + a_{49} u^2)
\; .
\end{array}
\end{equation}
Additionally $n_{bg}=3$ parameters are implemented to fit the
background
\begin{equation}
bg(x,y) = a_{50} + a_{51} x + a_{52} y \; .
\end{equation}

To get rid of the problem of a spatially varying PSF over the whole
area of the CCD, as described in \citet{1996AJ....112.2872T},
we divide the images in regions, each containing 141 x 141 pixels.
In each of these regions a locally valid convolution kernel is
calculated.
Since $K(u,v)$ has a box size of 21 x 21 pixels and the regions to be
convolved are overlapping, each region comprises 161 x 161 pixels. 
For all bad pixels (marked as 0) the convolution is not done,
these pixels remain 0.

The calculation of the matrix $M_{ij}$ is the most time consuming part
of the convolution.
The matrix $M_{ij}$ of the reference frame $R$ is calculated once
and can be used for all images. 
To enable this timesaving approach we take the error $\sigma_{x, y}$
which enters the calculations always from the error frame of $R$
($\sigma_{x, y} = \delta_R(x, y)$).
Therefore only the calculation of the vector $V_j$ has to be done
for each image/reference frame pair.

Bad pixels in the frame $I$ would lead to an error because they are
not marked in the frame $R$.
To compensate this the calculation of the matrix is redone for these
pixels and then subtracted from the original matrix.

After the convolution the difference frame $D$ is computed by
subtracting the $\tilde{R}$ frame from the $I$ frame
\begin{equation}
D = I - \tilde{R} \; .
\end{equation}
The calculation of the error frame is done according to Gaussian error
propagation
\begin{eqnarray}
\delta_{\tilde{R}}(x, y) & = &
\sqrt{\sum_{u, v} K^2_{u, v} \; \delta_{x+u, y+v}^2}
\quad \mbox{and}\\
\delta_D(x, y) & = &
\sqrt{\delta^2_{\tilde{R}}(x, y) + \delta^2_{I}(x,y)}
\;.
\end{eqnarray}
\subsection{Source detection}
\label{ss.detection}
We developed a standard star finding algorithm to detect 
sources in the difference images:
We smooth the image by replacing each pixel with the mean of five
pixels of a cross shaped area and tag all local maxima.
Then we fit a simplified Moffat function
\citep{1969A&A.....3..455M}
$A\left[ 1 + s (x^2+y^2)\right]^{-2.5}$
to these local maxima in the unsmoothed image.
After excluding all bad fits we fit a rotated Moffat function to the
remaining maxima:
\begin{equation}
\label{e.moffat}
f_\mathrm{moffat}(x,y)= A \left[ 1+(s_x x')^2+(s_y y')^2 \right]^{-\beta}
\;,
\end{equation}
where\\
\begin{tabular}{lcl}
$x'$& $=$ & $\left(x - x_0\right) \cos \alpha + \left(y - y_0\right)
\sin \alpha$,\\
$y'$& $=$ & $\left(y - y_0\right) \cos \alpha - \left(x - x_0\right)
\sin \alpha$.\\
\end{tabular}
\\
$\alpha$ denotes the rotation angle, $A$ the amplitude and the pair 
$x_0, y_0$ the central coordinates of a stellar PSF. 
The rise of the wings of the PSF is given by the parameter $\beta$,
whereas  $s$, $s_x$, and $s_y$ specify the width of the function.\\
We include the errors taken from the error frame in the nonlinear least
squares fit considering the error frame to weight the count rates
obtained in the frames.

Minimum and maximum expected FWHM of the
PSF and minimum and maximum of $\beta$
have to be chosen according to observational conditions.
To distinguish between noise and real sources a threshold factor
$t$, is introduced;
$t$ gives the ratio of the parameter $A$ and the background noise.
All sources below a certain threshold
(i.e. $t=5$ for the WeCAPP project)
are regarded as noise.
Because difference images can comprise negative sources the
images are inverted after one detection cycle.
The whole detection procedure is then redone on this inverted frame. 
\subsection{Photometry}
\label{ss.photometry}
Photometry of the detected sources is performed with a profile fitting 
technique.
We choose reference stars in the CCD
field to obtain the information about the PSF of any
particular frame.
These stars should be bright and isolated enough to allow an
accurate determination of the PSF.
On the other hand they have to be unsaturated in any of the images.
Since all frames are already photometrically aligned
(Sect.~\ref{ss.photalign}),
the amplitude of standard stars is not used for flux calibration,
so standard stars may even be variable.
We apply a Moffat fit
(Sect.~\ref{ss.detection}, Eq.~\ref{e.moffat})
to some standard stars selected in such a way in each image.
Keeping the slope parameters $\beta$, $s_x$, $s_y$, and the
angle $\alpha$ of the best fit star only the amplitude $A$ has to be
determined for each detected source in the difference frame by a
linear least-squares fit using propagated errors.
We integrate the count rates over the area of the now fully
known analytical function of the PSF of the source to determine its
flux.
This minimises the contamination from neighbour sources.

The fitting error of the amplitude
\begin{equation}
\delta_A =
\bar{\chi}_{\mathrm{source}} \;
\sqrt{ \frac{1}{\sum\limits_{x, y}^{n_\mathrm{fit}}
\frac{f_{\mathrm{moffat}}(x, y)}{A \; \delta^2_D(x, y)} }}
\end{equation}
is transformed into the
error of the flux determination by multiplying $\delta_A$ with the
flux corresponding to $A=1$. 
To account for the accuracy of the PSF-parameter determination we
multiply the resulting error with the $\bar{\chi}_{\mathrm{standard\;star}}$
(Eq.~\ref{e.chisquare}) of the standard star fit:
\begin{equation}
\delta_\mathrm{Flux} =
\delta_A \; \mathrm{Flux}_{A=1} \; \bar{\chi}_{\mathrm{standard\;star}}
\; .
\end{equation}
\section{Simulation tests with errors and error propagation}
\label{s.errors}
\tabcolsep1.2mm
To show the influences of a propagated
error estimation we have performed image reduction with and without
error propagation on simulated images, which include all the image
properties of Sect.~\ref{ss.ccd_properties}.
These simulations were done in addition to the empiric examination of
real data to have well defined input parameters and therefore be
able to decide whether the expected performance can be achieved.
All test images represent closely one of our detector configurations:
1k~$\times$~1k CCD, 16~Bit ADC, $\mathrm{gain}=3$, $
\mathrm{bias}=200\pm3~\mathrm{ADU}$,
$\mathrm{saturation}=65\,000~\mathrm{ADU}$.
\subsection{Errors in flatfields}
\label{ss.flat_errors}
We show the impact of error propagation by flatfielding artificial
skylight images with both flatfields and images representing different
flux levels.
Tab.~\ref{t.flat} gives the measured
relative errors i.e. the normalised standard deviation of the image.
The realistic $\sim30\mathrm{k}$ counts flatfields case compared to the
naive $\bar{\delta_I}$ case (Eq.~\ref{eq.bias} without any further
correction for flatfielding) gives an underestimation of $5\%$ to
$30\%$ compared to the true errors.
For the very low counts (but clean of stars and cosmics) case the
underestimation is even $50\%$.
The mean propagated error estimate, calculated as shown in
Sect.~\ref{ss.flatfield}, always differs less than
$2\%$ from the measured error.
\begin{table}
\centering
\small{
\begin{tabular}{cc|rrrrrrr}
\hline
\hline
\multicolumn{2}{c|}{average flux}&
\multicolumn{7}{c}{flat skylight image}\\
\multicolumn{2}{c|}{[$\times 10^3$ ADU]}&
1 & 10 & 20 & 30 & 40 & 50 & 60 \\
\hline
& $\overline{\delta}_I$ &
1.826 & 0.577 & 0.408 & 0.333 &
0.289 & 0.258 & 0.236 \\
f & 10 &
1.937 & 0.716 & 0.582 & 0.528 &
0.500 & 0.482 & 0.469\\
l & 20 &
1.916 & 0.656 & 0.507 & 0.445 &
0.411 & 0.389 & 0.373\\
a & 30 &
1.910 & 0.636 & 0.479 & 0.413 &
0.376 & 0.352 & 0.335\\
t &40 &
1.907 & 0.625 & 0.465 & 0.396 &
0.358 & 0.332 & 0.314\\
s & 50 &
1.905 & 0.619 & 0.456 & 0.386 &
0.346 & 0.320 & 0.301\\
& 60 &
1.903 & 0.614 & 0.450 & 0.379 &
0.338 & 0.312 & 0.292\\
& $\sim30$ &
1.911 & 0.640 & 0.485 & 0.420 &
0.382 & 0.359 & 0.343\\
\hline
\end{tabular}
\caption{\label{t.flat}
Normalised standard deviation of flatfielded artificial skylight
images [$\%$]:
$\overline{\delta}_I$ denotes a perfect flat, with noise
exclusively induced by the skylight image, i.e. the naive error;
rows 10 to 60 in respect to the $\overline{\delta}_I$ row illustrate
the impact on the error budget for different flux levels of the median
of five flatfield calibration image;
$\sim30$ shows the realistic case of five artificial dithered flats
comprising stars and cosmics.}}
\end{table}
\subsection{Errors with Gaussian filter for cosmic rays}
\label{ss.cosmic_errors}
Empirical tests on real images were done by comparing the number of
cosmics found in exposed images with that found in dark frames and
visually examining both the unfiltered image and the difference of the
unfiltered and the filtered image (e.g. for effects on bright stars
etc.).

In addition we have tested the reliability of our Gaussian
filter with five simulated cases
(Tab.~\ref{t.cosmicfiltertest}):
A pure skylight image, a simple field with 500 plus
$(N_\mathrm{sat}=)$ 10 saturated and bloomed stars and two different
sky levels, a crowded field with $100\,000$ stars, and a highly
crowded field with $200$~million stars; 
the positions of stars and cosmics as well as the flux and the
orientation of cosmics follow uniform deviates;
the flux of stars follows a exponential deviate,
which is a sufficient match to the luminosity function of our fields.
Stars have a PSF FWHM~$\approx$~2.6~pixel.
The images are processed with our standard reduction pipeline
(Sect.~\ref{s.standard}) using a median-of-five simulated flatfield
with an average flux level of $30\,000~\mathrm{ADU}$ per flatfield and
the filter parameters of Sect.~\ref{ss.cosmicfilter}
unless stated otherwise
(detection thresholds $t_\mathrm{limit}=8$,
$\mathrm{FWHM}_{\mathrm{cosmic}}=1.5$).

We determine the false alarm rate by filtering the clean test
images without any cosmics (Tab.~\ref{t.cosmicfilterperformance}):
For $10^6$ pixels with about 30 000 to 100 000 local maxima 0.3 to
$1 \times 10^5$ tests are performed.
So the false alarm rate is given by the ratio of false
occurances to number of tests.
To determine the detection rate we put 500 artificial cosmics
with flat deviates in space, form, and energy into the test images
(Tab.~\ref{t.cosmicfilterperformance}):
We identify and count the cosmics found.
To get an accurate estimate of the performance these numbers still
have to be compared with the photon noise, the noise induced by the
object density, and the filter parameters.

The highest false alarm rate is achieved with the crowded field.
Here the total pixel-to-pixel variation of the image
(photon noise plus objects' signal)
exceeds the pure photon noise by a factor of 25.
In the highly crowded field this excess is only a factor of 11 and
can be compensated by setting $t_\mathrm{limit}=10$.
The false alarm $\propto N_\mathrm{sat}$ in the simple, high sky field
is due to the unawareness of saturation\footnote{As
stated in Sect.~\ref{ss.cosmicfilter} saturated pixels need a special
treatment.}
because of missing saturation tags without error propagation.
Our tests show a false alarm rate $< 0.2\%$ in any (error propagated)
case.
We found that our fields are resembled closer by the smooth
simple-and-high-sky field than by the crowded and even the highly
crowded test fields.
Therefore we are sure the false alarm rate does not exceed $0.01\%$ in
our real images.
However, a false alarm resulting in a deletion of a true source
will not lead to wrong photometry because of the high error assigned
to replaced pixels
(Sect.~\ref{ss.cosmicfilter}).
Under certain circumstances (bad sampling, small stacks) it may lead
to large error bars, but the result is still reliable within those.

The expected detection rate ($97\%$ to $99\%$) is achieved nearly
with all test images including error propagation.
Even without error propagation the detection rate is still very close
to the expected one.
The worst case is again the (not highly) crowded field where the
object induced noise exceeds the photon noise by far.
Here the expected rate is missed by $0.008 = 4$ (of 500) cosmics.
\begin{table}
\centering
\small{
\begin{tabular}{c|r|rr|rr}
\hline
\hline
& sky & \multicolumn{2}{c|}{stars} & \multicolumn{2}{c}{cosmics} \\
\raisebox{1.5ex}[-1.5ex]{field} &
{level} &
number & max. flux & number & max. flux\\
\hline
empty & 500 & -- & -- & 500 & $80\,000$ \\
a) simple & & 500 & $30\,000$ & \\
+ low sky &
\raisebox{1.5ex}[-1.5ex]{500} & (+10) & ($200\,000$) &
\raisebox{1.5ex}[-1.5ex]{500} & \raisebox{1.5ex}[-1.5ex]{$80\,000$}\\
b) simple & & 500 & $30\,000$ \\
+ high sky &
\raisebox{1.5ex}[-1.5ex]{5000} & (+10) & ($200\,000$) &
\raisebox{1.5ex}[-1.5ex]{500} & \raisebox{1.5ex}[-1.5ex]{$80\,000$}\\
crowded & 500 & $100\,000$ & 1000 & 500 & $80\,000$ \\
high & 500 & $200 \cdot 10^6$ & 10 & 500 & $80\,000$ \\
\hline
\end{tabular}
}
\caption{\label{t.cosmicfiltertest}
Parameters of the test images for the Gaussian filter
for five different test configurations (fields) showing the level
of background sky [ADU], numbers and maximum fluxes [ADU]
of stars and cosmics.}
\end{table}
\begin{table}
\centering
\small{
\begin{tabular}{c|rrr|rrr}
\hline
\hline
error&&&&\\
propagation&
\multicolumn{3}{c|}{\raisebox{1.5ex}[-1.5ex]{with}}&
\multicolumn{3}{c}{\raisebox{1.5ex}[-1.5ex]{without}}\\
\hline
detection &
\centercell{false} & \multicolumn{2}{c|}{detection} &
\centercell{false} & \multicolumn{2}{c}{detection}\\
rates &
\centercell{alarm} & \multicolumn{2}{c|}{rate} &
\centercell{alarm} & \multicolumn{2}{c}{rate} \\
\hline
empty & $< 10^{-5}$ & 0.992 & \dag & $< 10^{-5}$ & 0.992 & \dag \\
a) simple & $< 10^{-5}$ & 0.992 & \dag & $\sim 10^{-5}$ & 0.988 & \\
b) simple & $\sim 10^{-5}$ & 0.982 & \dag &
$\propto N_\mathrm{sat}$ & 0.978 & \\
crowded & $1.80 \cdot 10^{-3}$ & 0.978 & &
$2.27 \cdot 10^{-3}$ & 0.972 & \\
high & $0.34 \cdot 10^{-3}$ & 0.976 & \dag &
$0.36 \cdot 10^{-3}$ & 0.964 & \\
high$^\mathrm{a}$ &
$\sim 10^{-5}$ & 0.970 & \dag &
$\sim 10^{-5}$ & 0.958 & \\
\hline
\end{tabular}
}
\caption{\label{t.cosmicfilterperformance}
Performance of Gaussian filter for cosmic ray events
for the test configurations of Tab.~\ref{t.cosmicfiltertest}:
The false alarm and the detection rates with and without error
propagation frames;
\dag~indicates that the detection rate
matches the expected rate according to
$t_\mathrm{limit}$ within 2 cosmics = $0.4\%$.
(a: $t_\mathrm{limit}=10$.)}
\end{table}
\subsection{Accuracy of the linear projection}
\begin{figure}
\centering
\includegraphics[width=0.5\textwidth]{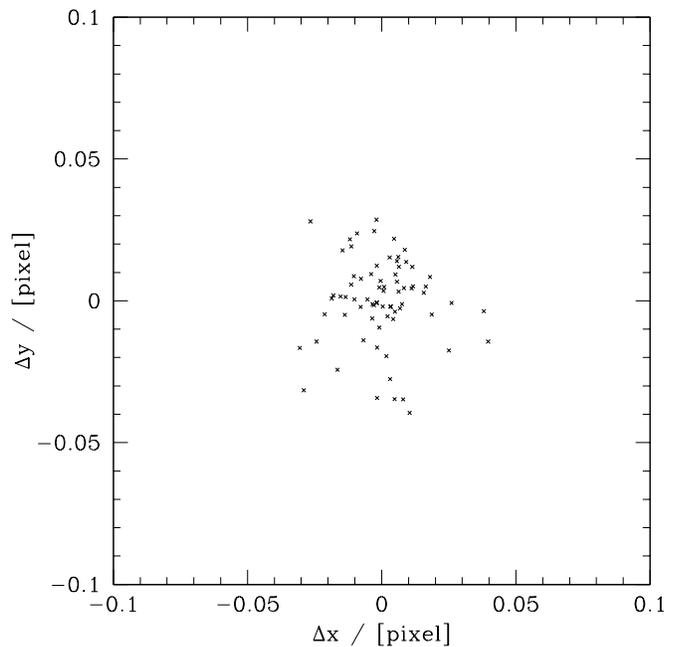}
\caption{\label{trans} Accuracy of the projection: 
position differences ($\Delta x=x_1'-x_2$, $\Delta y=y_1'-y_2$) of 70 stars
after projecting the coordinates of one frame ($x_1$,$y_1$) to the 
other frame ($x_2$,$y_2$).}
\end{figure}
The projection (Sect.~\ref{ss.projection}) 
was tested with two simulated, not perfectly 
aligned (shifted, rotated and rescaled) frames. 
It was calculated to
match the position of 70 bright stars in these frames.
The position differences are always below 0.05 pixels
(Fig.~\ref{trans}).
This reflects the principle accuracy limit\footnotemark due to the
size of the corresponding fit box
($20 \times 20$ pixels).
\footnotetext{Spatially
extremely undersampled images, leading to peak-shaped PSFs,
still can restrict this principle accuracy limit to one pixel,
but this was not a required test case.}
\subsection{Errors visible after alignment -- a snapshot}
\begin{figure*}
\centering
\begin{minipage}{\textwidth}
\includegraphics[width=0.485\textwidth]{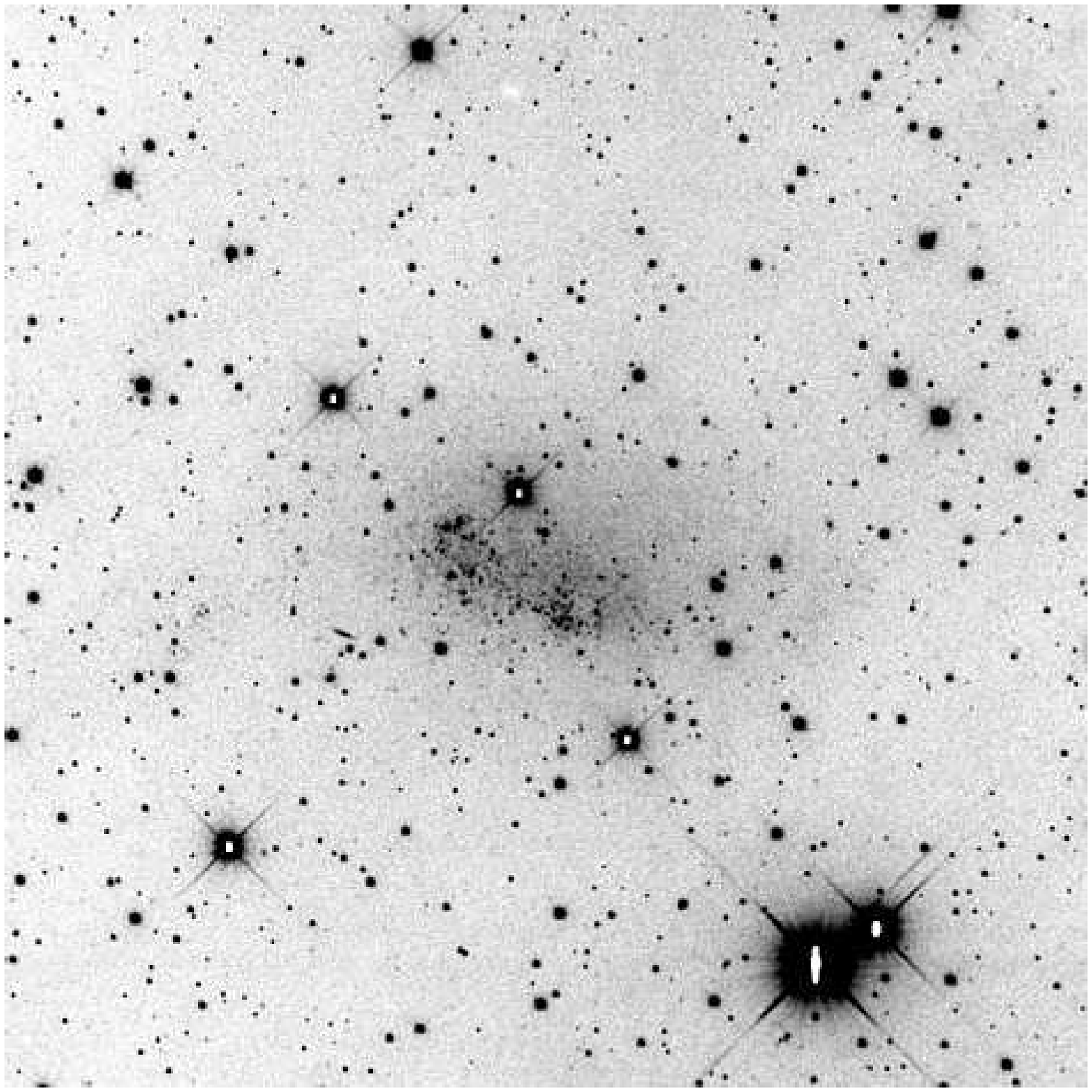}
\hspace*{\columnsep}
\includegraphics[width=0.485\textwidth]{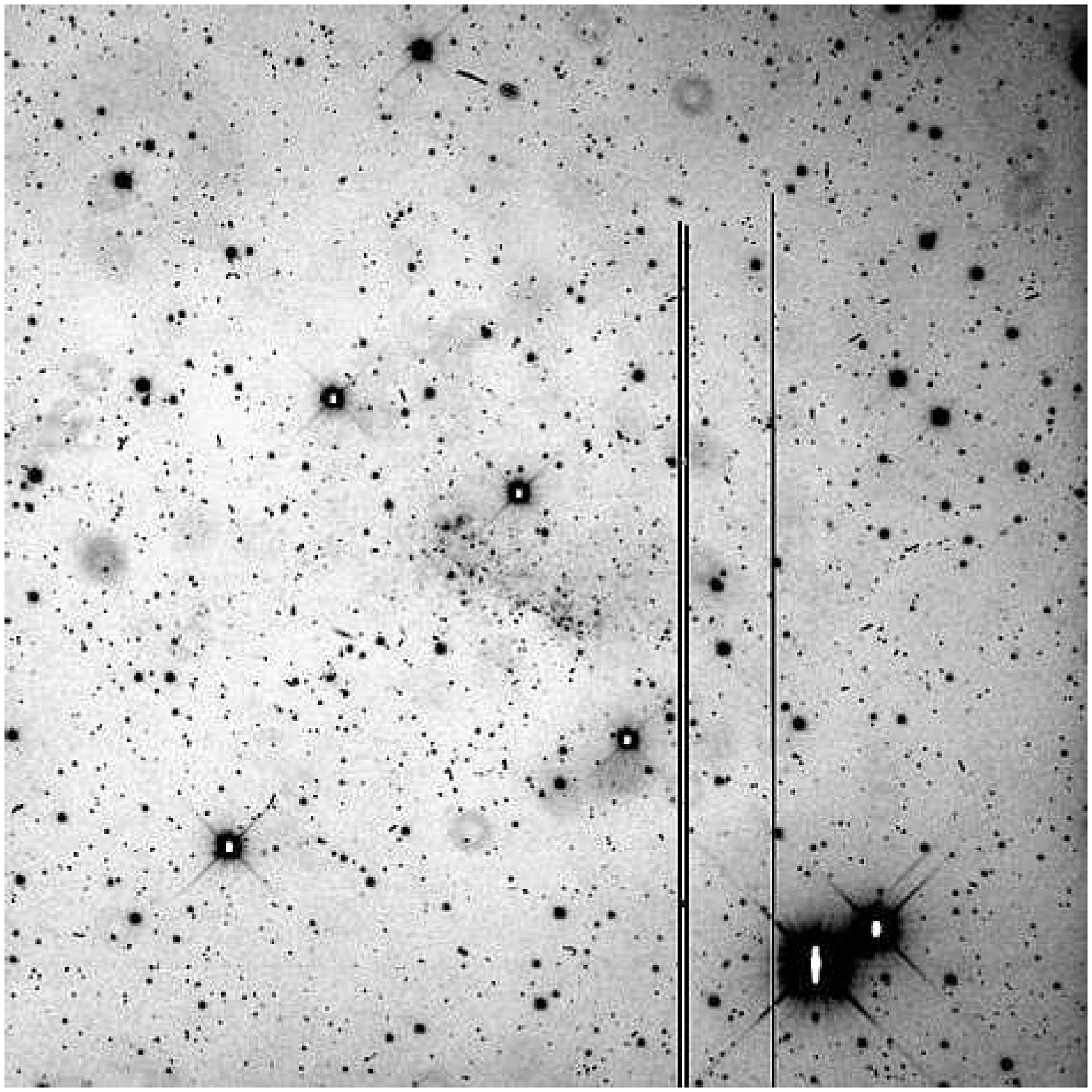}
\caption{\label{f.egb}The dwarf galaxy EGB\,0427\,+63:
One hour light on 0.8~m WST telescope,
best seeing and lowest sky 20 images of four days stacked;
left: image; right: error image.}
\end{minipage}
\end{figure*}

To give an impression of error features which would be neglected by
just considering the cleaned image,
but still visible in a propagated error image,
we present image and error image of one hour light on a small telescope
of the dwarf galaxy EGB\,0427\,+63
(Fig.~\ref{f.egb}).
Despite the fact that the images were dithered there are still
features of the flatfield (dust rings) clearly visible as well as the
impact of CCD defects and a huge amount of cosmics.
\subsection{Errors in convolution}
\begin{figure}
\centering
\includegraphics[width=0.5\textwidth]{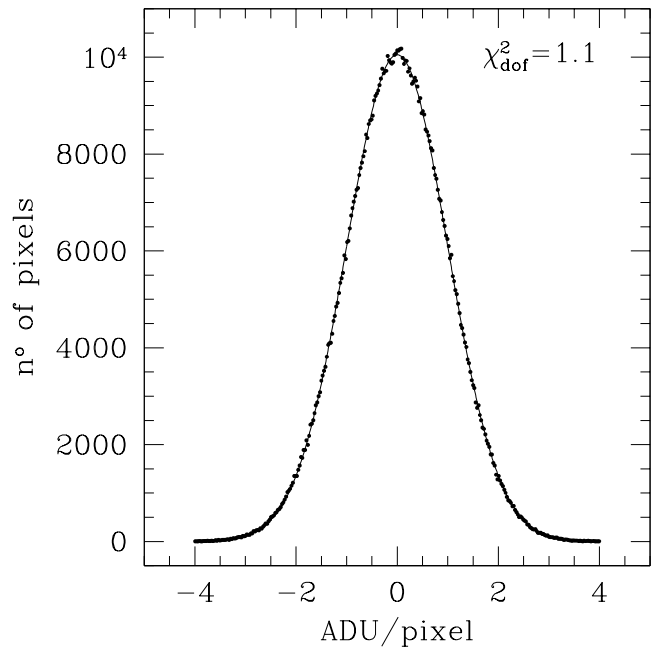}
\caption{\label{photon_noise}
Histogram of the pixel values of a simulated difference image divided
by the expected RMS errors.
The solid curve is a Gaussian with $\sigma=1$.
We calculate the reduced chi-square $\bar{\chi}^2$ of 19 different
simulated images in the range between -3 and 3.
The median is 1.1,
which means that expected and measured errors match almost perfectly
and that the residuals in the OIS are at the photon noise level.}
\end{figure}

We tested the accuracy of the convolution with 19 pairs of simulated
frames:
The reference frames with a FWHM of the stellar PSF of 2.4 and five
times more flux than the comparison frames with a FWHM of 3.0, all
with over 100~000 stars and different background levels. 
According to Sect.~\ref{ss.convolution} the reference frames are
convolved to match the PSF and the background level of the comparison
frames.
The convolved frames are subtracted from the comparison frames and the
result divided by the expected RMS errors, as derived from error
propagation.
This gives the ratio of expected photon noise and measured noise.
The histogram of such a ratio frame matches a Gaussian with $\sigma=1$
almost perfectly, which means that the expected photon noise fits the
measured noise.
This shows that the OIS method can be applied to very crowded fields
like M\,31 and gives residual errors at the photon noise level 
(Fig. \ref{photon_noise}).
\subsection{Errors in interpolations -- PSF-fitting photometry}
\label{ss.interpolation_errors}
We performed PSF-fitting photometry in simulated images as described
in Sect.~\ref{ss.photometry} using a Moffat fitting function
(Eq.~\ref{e.moffat})
on bright stars, and
(together with OIS, using a high signal-to-noise reference image)
on faint variable sources in a crowded and a highly crowded field.
We found that per pixel propagated errors compared to estimated
errors greatly enhanced the reliability of any fit.
Extensive testing with simulated images comprising different
observational features
shows that this is especially due to the treatment of defective
pixels, cosmics and saturation (pixel defects).
If we want to avoid the labour of full error propagation
we nevertheless have to use masks to get rid of these pixel defects.
But then long time series will diminish our field, because the defects
will spread
(due to dithering, minor misspointing, different detectors, random
position defects etc.).
Furthermore,
the $\bar\chi^2$ (Eq.~\ref{e.chisquare}) of a fit has a valid
meaning only for fully propagated errors:
$\bar\chi^2 \approx 1$ implies a correct measurement within purely
noise induced errors and correctly flagged pixel defects;
$\bar\chi^2 > 1$ indicates systematic errors beyond noise
and corrected pixel defects like blending of variable sources, missed or
wrongly treated pixel defects.
Renouncing full error propagation will shift e.g. bad flatfields from
the recognised, high noise category to the undiscovered systematics
regime.
The calculated error will not change, because we consider
$\bar\chi^2$ for the final error budget,
but we would lose the chance to investigate those cases.
If one cannot prove the origin of the uncertainties in a
measurement,
one can neither be sure of the measurement itself nor of the
postulated accuracy.
\section{Summary}
\label{s.summary}
We have presented a standard image reduction pipeline
applicable to any CCD images,
but with the intention to execute DIA,
i.e. image convolution and relative photo\-metry.
We put a great emphasis on the importance of per pixel propagated
errors.
The necessity of building good flatfield calibration images and a way
to obtain them was shown.
We also presented a robust filtering technique for cosmic rays
applicable to single, not too undersampled images.
We discussed all image reduction issues of
finding variable objects and measuring their variations in highly
crowded fields:
PSF matching by convolution of a reference image using the
\citet{1998ApJ...503..325A} technique,
construction of difference images,
detection algorithm for variable sources,
and relative photometry of variable sources by a profile fitting
technique.

A paper about the application of this image reduction pipeline on a
massive imaging campaign of a part of the M\,31 Bulge is accepted by
A\&A \citep{2001A&A...0000..00R}.
Parts of the reduction pipeline have also been successfully applied
to MUNICS data
(cosmic-filtering of MOSCA spectra and CAFOS images;
\citealp{2001MNRAS.325..550D}) and
VLT FORS data (cosmic-filtering and image alignment of revised FDF
frames, A.~Gabasch, priv.\ comm.;
image alignment and OIS
in the centre part of NGC\,4697 using a difference image built
of narrow on-band and off-band line images,
\citealp{2001ApJ...0000..00R}).
\begin{acknowledgements}
Our thanks are due to R.~Bender, U.~Hopp, and R.~P.~Saglia for giving us
a start and many hints in image reduction,
to A.~Fiedler, N.~Drory, and J.~Snigula for invaluable insights on the
C++ programming language,
and, together with G.~Feulner, for help with the A\&A \LaTeX\  document
class and  {\sc Bib}{\TeX},
and to U.~Hopp, R.~P.~Saglia, and N.~Drory, again,
for their review of the manuscript.
We also thank the referee Dr. R.~Butler for his kind and
encouraging report and for his help in improving the manuscript.
This work was supported by the German \emph{Deut\-sche
For\-schungs\-ge\-mein\-schaft, DFG}, SFB 375
Astro\-teil\-chen\-phy\-sik.
\end{acknowledgements}
\end{document}